\documentclass[12pt]{iopart}
\usepackage{graphicx}
\usepackage{amssymb}
\usepackage{color}
\usepackage{amstext}
\usepackage{latexsym}
\usepackage[colorlinks,linkcolor=blue]{hyperref}

\newcommand{\beqa}{\begin{eqnarray}}
\newcommand{\eeqa}{\end{eqnarray}}

\begin{document}

\title{Topological relics of symmetry breaking: Winding numbers and scaling tilts from random vortex-antivortex pairs.}
\author{W. H. Zurek$^1$}
\address{$^1$Theory Division,  Los Alamos National Laboratory, Los Alamos, NM, 87545, USA}

\begin{abstract}
I show that random distributions of {\it vortex-antivortex pairs} (rather than of {\it individual vortices}) lead to scaling of typical winding numbers ${\cal W} $ trapped inside a loop of circumference ${\cal C}$ with the square root of that circumference, ${\cal W} \sim \sqrt {\cal C}$, when the expected winding numbers are large, $|{\cal W}| \gg 1$. Such scaling is consistent with the Kibble-Zurek mechanism (KZM), with  ${\langle {\cal W }^2 \rangle}$ inversely proportional to $ {\hat \xi}$, the typical size of the domain that can break symmetry in unison. (Dependence of $ {\hat \xi}$ on quench rate is predicted by KZM from critical exponents of the phase transition.)
Thus, according to KZM, dispersion $\sqrt{\langle {\cal W }^2 \rangle}$ scales as $\sqrt { {\cal C} / {\hat \xi}}$ for large $\cal W$.
By contrast, distribution of individual vortices with randomly assigned topological charges would result in dispersion 
scaling with square root of the area inside ${\cal C}$ (i.e., $\sqrt{\langle {\cal W }^2 \rangle} \sim {\cal C}$). 

Scaling of the dispersion of $\cal W$ as well as of the probability of detection of non-zero $\cal W$ with $\cal C$ and $ \hat \xi$ can be also studied for loops so small that non-zero windings are rare. In this case I show that dispersion varies not as $1/ \sqrt{{\hat \xi}}$, but as $1/{\hat \xi}$, which results in a {\it doubling} of the scaling of dispersion with the quench rate when compared to the large $|{\cal W}|$ regime. Moreover, the probability of trapping of non-zero $\cal W$ becomes approximately equal to ${\langle {\cal W }^2 \rangle}$, and scales as $1/{\hat \xi}^2$. This {\it quadruples} -- as compared with $\sqrt{\langle {\cal W }^2 \rangle} \simeq \sqrt { {\cal C} / {\hat \xi}}$  valid for large $\cal W$ -- the exponent in the power law dependence of the frequency of trapping of $|{\cal W}|=1$ on ${\hat \xi}$ when the probability of $|{\cal W}| > 1$ is negligible. 
This change of the power law exponent 
by a factor of {\it four} -- from $1/\sqrt {\hat \xi}$ for the dispersion of large $\cal W$ to $1/{\hat \xi}^2$ for the frequency of nonzero $\cal W$ when $|{\cal W}| > 1$ is negligibly rare -- 
is of paramount importance for experimental tests of KZM.
\end{abstract}

\pacs{64.60.Ht, 05.30.Rt, 73.43.Nq}

\maketitle

\section{Introduction}

Formation of topological defects in symmetry-breaking phase transitions \cite{Kibble76,Kibble80,Kibble07,Zurek85,Zurek93,Zurek96} yields a seemingly random distribution of, for example, vortices in a superfluid or flux lines in a superconductor. Yet, according to KZM, this randomness comes ultimately from the local, uncorrelated choices of the broken symmetry -- e.g., choices of the phase of the condensate wavefunction. This also means that topological charges of the neighbouring defects are anticorrelated, as noted some time ago by Liu and Mazenko \cite{LM} and, more recently, seen in the experiments involving superconducting thin films \cite{Golubchik}. 


This primacy of randomness of the local choice of broken symmetry over 
the defect charge 
is reflected in typical winding numbers $\cal W$ subtended by a loop of circumference ${\cal C}=2 \pi R$ immersed in the superfluid or placed on the surface of the superconducting film. If charges of the vortices intercepting 2-D plane defined by the loop were random -- i.e., completely uncorrelated -- typical $\cal W$ would scale with the square root of the total number of vortices within $\cal C$, so it should be proportional to the square root of the area of the loop, $\pi R^2$, or, in other words, directly proportional to its circumference $\cal C$. 

By contrast, when phase of the superfluid wavefunction is random, typical winding number is proportional to square root of ${\cal C} / {\hat \xi}$, the number of domains of typical size $\hat \xi$ -- regions small enough to break symmetry in unison -- through which $\cal C$ passes \cite{Zurek85}. That is, in the case of random phases (local choices of broken symmetry implied by KZM):
\beqa
\sqrt{\langle {\cal W}^2 \rangle } \sim \sqrt {\frac {\cal C} {\hat \xi}} \sim \sqrt {\frac R {\hat \xi }}\ .
\eeqa
Dispersion yields estimates of the typical winding numbers. This scaling is consistent with simulations \cite{DSZ12} and experiments \cite{Maniv03,Maniv05}, and can be derived for  transitions from an $O(N)$ symmetric state to a phase of broken symmetry for large $N$ \cite{Fischer1,Fischer2,Fischer3}.

This paper points out that 
while the naive picture of a random distribution of vortex charges on the 2-D plane is inconsistent with KZM, a simple (and almost as naive) model where, instead, vortex-antivortex {\it pairs} are randomly distributed on a plane accounts (at least qualitatively) for the anticorrelation and, above all, predicts scalings of the typical winding numbers subtended by a loop consistent with Eq. (1) -- with KZM. 

We shall also explore scalings expected for $\langle |{\cal W}| \rangle \ll 1$ (so that frequency with which ${\cal W}=\pm 1$ is trapped becomes the focus of attention). For this case we arrive at results that shed new light on the interpretation of experiments and that are likely valid outside of the range of the random vortex-antivortex pair model. For $\langle |{\cal W}| \rangle \ll 1$, non-zero $\cal W$ occurs almost exclusively when $\cal C$ contains a single defect. Thus, ${\cal W}=\pm 1$  is given by the probability of finding a defect inside $\cal C$, which, for a contour with an area $A_{\cal C}$ is:
$$
p_{{\cal W}=\pm 1} \sim \frac {A_{\cal C}} {\hat \xi^2} \sim \biggl( {\frac R {\hat \xi } } \biggr)^2  \eqno(1')
$$
when, in accord KZM, we take the (surface) density of defects to be $1/\hat \xi^2$.
This power law dependence on $\hat \xi$ involves exponent four time larger than the dispersion for large typical $\cal W$, Eq. (1), which means a four times steeper dependence of the quench rate.

\section{Choosing phases in the Kibble-Zurek mechanism}

Emergence of broken symmetry phase in systems such as superfluids and superconductors involves local choices of broken symmetry ``vacua''. It is now well understood that such local choices of broken symmetry can lead to creation of topological defects via the Kibble-Zurek mechanism (or KZM) \cite{Kibble76,Kibble80,Zurek85,Zurek93}. The first basic observation is that causally disconnected fragments of the new phase must have acted independently \cite{Kibble76,Kibble80}. Moreover, in the second order phase transitions the relaxation time and the healing length scale as:
\beqa
\tau(t)  =  \frac{\tau_{0}}{|\varepsilon|^{z\nu}}, \ \ \ \ \ \xi(t)  =  \frac{\xi_{0}}{|\varepsilon|^{\nu}},
\eeqa
where $\nu$ and $z$ are the critical exponents that define the universality class of the transition and the relative distance from the critical point is quantified by the dimensionless:
\beqa
\varepsilon=\frac{\lambda_c-\lambda}{\lambda_c}.
\eeqa
In the situation of interest to us $\varepsilon$ changes with time, driving the system though its critical point. For instance, $\lambda$ could stand for temperature (and $\lambda_c$ could be then the critical temperature) and phase transition could be induced by cooling. Or, for the case of quantum phase transitions, $\lambda$ could be a parameter of the Hamiltonian that induces quantum phase transition and $\lambda_c$ its critical value. The second basic observation is that -- because of the critical slowing down, i.e., the divergence of the relaxation time in the vicinity of the critical point -- the system cannot remain in equilibrium while undergoing a second order phase transition at any finite rate \cite{Zurek85,Zurek93}. 

In second order phase transitions the size of domains over which the system can coordinate the choice of the broken symmetry is decided by the scaling of the relaxation time with $\varepsilon$. When it is assumed that:
\beqa
\varepsilon=\frac{t}{\tau_Q}
\eeqa
one can show \cite{Zurek85,Zurek93} that the broken symmetry choice is associated with the instant:
\beqa
\hat{t}\sim\left(\tau_{0}\tau_{Q}^{z\nu}\right)^{\frac{1}{1+z\nu}} \label{eq:freeze_out_time}
\eeqa
after the critical point has been crossed at $t=0$. The corresponding size of the domains that can agree on how to break symmetry is given by:
\beqa
\hat{\xi} \simeq \xi(\hat{t})=\xi_{0}\left(\frac{\tau_{Q}}{\tau_{0}}\right)^{\frac{\nu}{1+z\nu}}.
\label{eq:correlation_length}
\eeqa
Domains of this size can still coordinate selection of the broken symmetry, but regions separated by more than $\sim\hat{\xi}$ must choose broken symmetry ``on their own'' \cite{Zurek85,Zurek93}, presumably with the help of the local (thermal or quantum) fluctuations.

As noted by Kibble  \cite{Kibble76} in the context of the field-theoretic symmetry breaking in the early Universe (where the domain size was bounded from above by relativistic causality) whenever homotopy group permits, approximately one topological defect is expected to form in a volume of a domain that can coordinate the choice of symmetry breaking. 

In view of the above discussion, in the phase transitions accessible in the laboratory such domains should have size of $\sim\hat{\xi}$ which then -- along with Kibble's observation cited above -- leads to the defect density estimates  \cite{Zurek85,Zurek93}. In the 2-D case with vortices piercing the plane this leads to the expected vortex density:
\beqa
n \sim {\hat{\xi}^{-2}} \ .
\eeqa

We focus on two dimensional case because it is easy to think about and to illustrate on a 2-D page, but also because it is experimentally relevant: Some of the earliest experiments testing the basic prediction of KZM -- formation of defects -- were carried out in 2-D \cite{Golubchik,Carmi99,Maniv03,Maniv05}. Indeed, early claim of results at odds with KZM was made on the basis of assumption of a random distribution of vortices in the resulting planar superconductor \cite{Carmi99}. It arose from the expectation that vortices (that are randomly scattered on the plane) have their topological charges selected also at random. If that was the case (as was noted already earlier) in a loop
that contains $n$ vortices one would expect an imbalance of a random sign but a typical magnitude $\Delta n = | { n}_+ - {n}_- | \sim \sqrt { n}$. This would lead one to expect a winding number $|{\cal W}| \sim \sqrt {n}$. Therefore, in a superconductor, a corresponding magnetic field flux proportional to the square root of the area subtended by $\cal C$, i.e. directly proportional to the circumference, $\cal W \sim \cal C$, would be predicted from the random arrangement of flux tube charges, at odds with KZM. 

For a random distribution of the wavefunction phases (which, according to KZM, captures the fundamental randomness) the scaling -- Eq. (1) -- is different. This conflict between the two scalings can be resolved when topological charges of neighbouring defects are anticorrelated, so that the net charge is smaller than predicted by the random distribution of charges. This is consistent with the analytic estimates of the correlation functions between defects created by a transition, Ref. \cite{LM}, and is also consistent with the data obtained in the 2D superconductor experiments.\footnote{We note that this expectation of an anticorrelation of defect charges ignores influence of the magnetic field -- or, more generally, gauge fields -- on the distribution of topological defects. This is clearly not always correct \cite{HindmarshRajantie,SBZ}, although it appears to be a good approximation in the superconducting thin film experiments of the Haifa group \cite{Golubchik}.}

\section{Winding number from vortex-antivortex pairs}

\begin{figure}
\begin{center}
\includegraphics[width=1.1\linewidth]{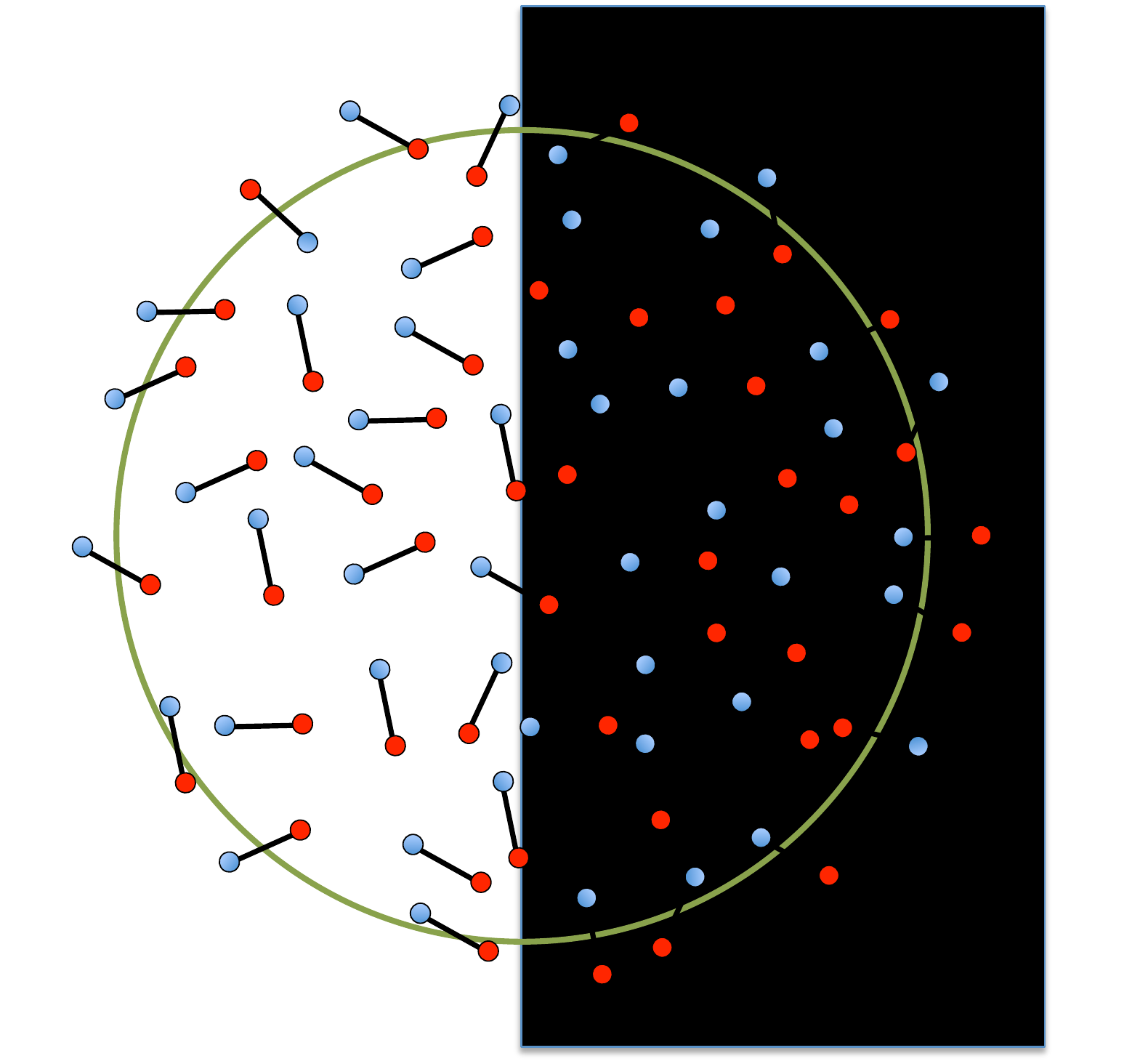}
\end{center}
\caption{\label{kzmpairs} We consider winding number along the circumference $\cal C$ (the circle above) due to vortex-antivortex pairs scattered randomly on a plane. Contribution to the winding number due to the pairs that are completely outside or completely inside the contour  $\cal C$ above vanishes. Thus, only pairs that straddle the contour will contribute. The number of such pairs is proportional to the circumference $\cal C$. When (as we assume) they have random orientations, typical winding numbers will be proportional to the square root of the number of pairs intercepted by $\cal C$, i.e., $ {\cal W} \sim \sqrt {\cal C} $, Eq. (1). In the case of KZM the typical separation of a vortex from an antivortex would be presumably $\sim \hat\xi$, and of the same order as the separation between the pairs. Note that pairing discussed in this paper is in a sense imaginary (as is suggested by the right hand side of the figure), as there is generally no unique ``correct'' way to combine vortices and antivortices into pairs. Nevertheless, recognition of pairing leads to correct scaling of winding numbers with $\cal C$ while accounting for the anticorrelation of the topological charges of defects -- the key physical implications that follow from the random choices of symmetry breaking predicted by KZM. When loops are so small that typically, at most only one end of a pair ``fits inside'', scaling changes, see Eqs. (8)-(11) and Fig. 2.}
\end{figure}

Our basic observation (see Fig. 1) is that when -- instead of individual vortices -- vortex pairs with the typical size (i.e., vortex-antivortex separation) $l$ small compared to $\cal C$ are randomly distributed on a plane, winding number is determined by the pairs that are cut by $\cal C$. Pairs that are completely outside or completely inside $\cal C$ do not count: Only pairs bisected by $\cal C$ contribute to $\cal W$:  
The number $n_{\cal C}$ of pairs cut by the circumference $\cal C$ is proportional to $\cal C$. When their orientations are random (as we assume), the difference between the topological charges will be given by the square root of the number of bisected  pairs, i.e., $\sqrt n_{\cal C}$. This reasoning (and Fig. 1) conclude our ``proof'' of Eq. (1) in the case of the random distribution of vortex-antivortex pairs.

The quotation marks immediately above are clearly deliberate: We have not even stated our assumptions as carefully as a theorem would demand, so our discussion is suggestive, but falls short of a real  ``proof''. 
Let us however point out a simple case in which the above assertions are provably correct: 
Consider a circle $\cal C$ with a radius $R$ much larger than the pair size $l$, so that one obvious complication (e.g., connecting line of a pair intersecting the contour $\cal C$ twice) can be neglected. 
The number of pairs that straddle $\cal C$ is proportional to the area of the strip defined by
the product of its circumference and the typical size of the pair $l$, and pair density given by the inverse of the area $L^2$ defined by the inter-pair separation $L$. 

When pair orientations are random, $\sim {\cal C} l / L^2$ pairs straddle the contour, so the typical net winding number will be
$$\langle {\cal W}^2 \rangle|_{{R} \gg l} \sim {\frac {{\cal C}l} {L^2}} \  \eqno(8a)$$
consistent with Eq. (1). Moreover, the
contour $\cal C$ need not be circular: As long as $\cal C$ has a topology of a circle (rather than, e.g., of a figure ``8'') and a curvature radius large compared to pair sizes as well as no features such as a ``peninsula'' or an  ``isthmus'' with width comparable to $l$ -- e.g., it is not an dumbbell with the ``handle'' thinner than pair size -- the above conclusion should hold. This means that the winding number -- for large loops, and given the above assumptions -- does not depend on the area inside $\cal C$, but only on the circumference.\footnote{Contours with large circumference ${\cal C} \gg l$ but relatively small area (e.g., $A_{\cal C} \ll l^2$) -- long and thin, e.g., spaghetti or hydra shaped contours -- may lead to complications. We shall not address them here. 
}

\section{Scaling tilts for small loops}

When the size of the contour is (in conflict with what was assumed above) small compared to $l$ (so that, for example, radius $R$ of circular $\cal C$ is small compared to $l$), then $\cal C$ can typically accommodate, at most, only one defect -- one end of the vortex-antivortex pair -- and the winding number is then either ${\cal W}= 0$ (``empty'' $\cal C$) or, otherwise, $|{\cal W}| = 1$ (while the frequency of cases with $|{\cal W}| > 1$ vanishes). The nature of the question that leads to the typical $\cal W$ changes from ``How many pairs straddle $\cal C$?'' (relevant when $R \gg l$) to ``Is there a single defect inside $\cal C$?'' (relevant when $R \ll l$). 

The probability of finding a defect inside $\cal C$ is $ {A_{\cal C}} / {L^{2}}$, the product of the planar density $L^{-2}$ of defects and the area $A_{\cal C}$, so the dispersion of the winding numbers found inside the contour of size $\cal C$ is:
$$\langle {\cal W}^2\rangle|_{{R} \ll l} \simeq (1- \frac {A_{\cal C}} {L^2})\times 0^2 +\frac {A_{\cal C}} {2L^2} \times 1^2 + \frac {A_{\cal C}} {2L^2} \times (-1)^2= \frac {A_{\cal C}} {{\hat \xi}^2}   = {\frac {\pi R^2} {L^2}} \eqno(8b)$$
where we have counted separately contributions from defects (${\cal W}=1$) and antidefects (${\cal W}=-1$) assuming that each has the density of ${ \frac 1  2 L^{-2}}$, and where the last equality assumes explicitly that $\cal C$ is circular.
In actual phase transitions both pair size and separations between pairs will likely scale with the size of domains where broken symmetry can be coordinated, $l \simeq L \simeq \hat \xi$. Using this in Eqs. ($8a,b$) we conclude that scaling of $\cal W$'s will have different dependence on the critical exponents in the two regimes. Thus, for $ \langle{\cal W}^2\rangle \gg 1 $, Eq. ($8a$) leads to;
$$ \langle{\cal W}^2\rangle|_{ \langle{\cal W}^2\rangle \gg 1} \sim {\frac {{\cal C}} {\hat \xi}} \sim  {\hat \xi}^{-1} \sim\left(\frac{\tau_{0}}{\tau_{Q}}\right)^{\frac{\nu}{1+z\nu}} \ , \eqno(9a)$$
while in the opposite case of small $ \langle{\cal W}^2\rangle$, Eq. (8b), implies:
$$ \langle{\cal W}^2\rangle|_{ \langle{\cal W}^2\rangle \ll 1} \sim \frac {A_{\cal C}} {{\hat \xi}^2} \sim \hat \xi^{-2} \sim\left(\frac{\tau_{0}}{\tau_{Q}}\right)^{\frac{2 \nu}{1+z\nu}} \eqno(9b)$$
In other words, the exponent in the dependence of dispersion (or of its square, computed above) of the winding numbers on the quench time $\tau_Q$ {\it doubles} between the two regimes.\footnote{Doubling of the power law for loops small enough so that
$\langle |{ \cal W} | \rangle<1$ (based on the case of tunnel Josephson junctions) was predicted some time ago \cite{KMR,RM1}.
This prediction of doubling was recently reevaluated and expanded \cite{NZ,RM2} to include exponential (rather than power law) dependence on $\tau_Q$ for very slow quenches that yield small (although not necessarily $|{\cal W}| < 1$) winding numbers (see \cite{DTZ} for other slow quench possibilities). We also note that numerical simulations \cite{Ueda,DzMZ} show evidence of steepening slope of the dispersion -- consistent with doubling -- for winding numbers trapped in BEC annuli when $ \langle{\cal W}^2\rangle \le 1 $. 
Our study is not focused on Josephson Junctions or superconducting loops. Still, our conclusions are based on statistics, and, therefore, may be broadly valid. One can think that a loop (represented by $\cal C$) that becomes superconducting  ``lassoes'' flux lines arising from random configurations of the order parameter inside $\cal C$, thus trapping $\cal W$.
}

There is one more question one could ask and that is, indeed, often asked in KZM experiments involving small systems.
Suppose that instead of dispersion (computed in Eqs. (8) and (9)) we estimate frequency of non-zero $\cal W$ or average absolute value of $\cal W$. (Probability of $|{\cal W}|\neq 0$ becomes equal to $\langle |{\cal W}| \rangle $ when the only possible values of $\cal W$ are +1, 0, and -1.)

In the range where $\langle{\cal W}^2\rangle \gg 1$ typical winding number will have its value set by dispersion, so that -- in accord with Eq. (1) -- $\langle |{ \cal W} | \rangle|_{ \langle{\cal W}^2\rangle \gg 1} \sim \sqrt { \langle{\cal W}^2\rangle}$, or, more precisely:
$$ \sqrt {\frac \pi 2}\langle |{ \cal W} | \rangle|_{ \langle{\cal W}^2\rangle \gg 1} =  \sqrt { \langle{\cal W}^2\rangle} \sim \sqrt {{\frac {{\cal C}l} {L^2}}} \sim {\hat \xi}^{-{\frac 1 2}} \sim \left(\frac{\tau_{0}}{\tau_{Q}}\right)^{\frac{\nu}{2(1+z\nu)}} \eqno(10a)$$
where the factor $\sqrt {\frac \pi 2}$ follows from Gaussian approximation to the distribution of $\cal W$'s (that -- as numerical simulations confirm \cite{DSZ12} -- is quite good also for moderate values of $\cal W$, down almost to $\langle |{\cal W} | \rangle \sim 1$). 

By comparison, when $\langle{\cal W}^2\rangle \ll 1$, Gaussian approximation to the distribution of $\cal W$'s breaks down, and $\langle | {\cal W} | \rangle $ as well as the probability $p_{{\cal W}\neq 0}\approx p_{{\cal W}=\pm1}$ of detecting a non-zero winding number (and, hence, the frequency of such detections) is given by:
$$ \langle |{ \cal W} | \rangle|_{ \langle{\cal W}^2\rangle \ll 1} \simeq p_{{\cal W}\neq 0} \simeq \frac { A_{\cal C}} {{\hat \xi}^2} \sim {\hat \xi}^{-2} \sim  \left(\frac{\tau_{0}}{\tau_{Q}}\right)^{\frac{2 \nu}{1+z\nu}} \eqno(10b)$$
Thus, in this case the scaling {\it quadruples} compared to that of Eq. (10a). 

It is important to pay attention to these differences in interpreting either numerical or experimental data, as dealing with different (if closely related) quantities such as $ \langle |{ \cal W} | \rangle$ and $ \sqrt {\langle { \cal W}^2 \rangle} $ or frequency (i.e., $p_{{\cal W}\neq 0}\approx p_{{\cal W}=\pm1}$) can lead to confusion.

The presence of two (re-)scalings -- doubling (for $ \sqrt { \langle{\cal W}^2\rangle}$) or quadrupling (for $\langle |{ \cal W} | \rangle$ or $p_{{\cal W}\neq 0}$) of the power law in the transition between the regime of large and small winding numbers has not been pointed out before, and, at first sight, may appear surprising. After all, we are accustomed to the approximate equality (first part of Eq. ($10a$)) of typical values and dispersions, so that one might expect $\langle |{ \cal W} | \rangle \sim  \sqrt { \langle{\cal W}^2\rangle}$. This proportionality should hold when the distribution of outcomes is approximately Gaussian. In our case it is a reliable approximation only when ${ \langle{\cal W}^2\rangle \gg 1}$. Then the scaling of $\langle |{ \cal W} | \rangle $ and of $\sqrt { \langle{\cal W}^2\rangle}$ indeed coincide, as both quantities are dominated by the magnitude of the typical values of $\cal W$. 

By contrast, when ${ \langle{\cal W}^2\rangle \ll 1}$, typical value of the winding number is zero, ${\cal W}=0$, with only rare (probability $p_{{\cal W}\neq 0}\simeq {A_{\cal C}} / {{\hat \xi}^2} \ll 1 $) exceptions.
As we have seen in Eqs. ($9b$) and ($10b$),  when ${ \langle{\cal W}^2\rangle \ll 1}$, both ${ \langle{\cal W}^2\rangle}$ and $\langle |{ \cal W} | \rangle$ are, to the leading order, given by ${A_{\cal C}} / {{\hat \xi}^2} $, the probability of trapping a single defect inside $\cal C$:
$${ \langle{\cal W}^2\rangle}|_{ \langle{\cal W}^2\rangle \ll 1} \approx p_{{\cal W}\neq 0}
\approx p_{|{\cal W}|=1}
\approx \langle 
|{ \cal W} | \rangle|_{ \langle{\cal W}^2\rangle \ll 1}\ . \eqno(11)$$
This sequence of approximate equalities becomes essentially exact as the probability of $|{\cal W}| > 1$ becomes negligible. 

This asymptotic coincidence, Eq. (11), of the quantities that can be otherwise (e.g.,  when $|{\cal W}| \ge 1$) quite different is easy to understand: In the limit of small typical winding numbers both the square of the dispersion ${ \langle{\cal W}^2\rangle}|_{ \langle{\cal W}^2\rangle \ll 1}$
and the absolute value $ \langle |{ \cal W} | \rangle|_{ \langle{\cal W}^2\rangle \ll 1}$ are given by the probability $p_{{\cal W}\neq 0}|_{ \langle{\cal W}^2\rangle \ll 1} \approx p_{{\cal W}=\pm1}|_{ \langle{\cal W}^2\rangle \ll 1}$ when the probabilities of higher values of $|{\cal W}|$ can be neglected. 

\section {Scaling tilts, small loops, and experiments}

There are important consequences of the above discussion that are relevant for experimental investigations (which often take place in the small $\cal W$ limit, and seek to determine scaling of the frequency of non-zero detections of the winding number as a function of the quench rate): While $\langle |{ \cal W} | \rangle$ scales as probability of non-zero winding number, $p_{{\cal W}\neq 0} \sim 1 /{{\hat \xi}^{2}}$, the dispersion $\sqrt{ \langle{\cal W}^2\rangle}$ scales as $ \sqrt { p_{{\cal W}\neq 0} } \sim 1/\hat \xi$, and $\hat \xi$ depends on the quench rate $1/\tau_Q$ as indicated in Eq. (6). Therefore, the power law dependence on the quench rate tilts -- it becomes steeper -- as the exponent increases by a factor of {\it four} between the dispersion of large typical winding numbers, $\sqrt{ \langle{\cal W}^2\rangle}|_{ \langle{\cal W}^2\rangle \gg 1}$ and the frequency of detections of winding $p_{{\cal W}\neq 0}|_{ \langle{\cal W}^2\rangle \ll 1}$ in the opposite limit (see Fig. 2). 

Such quadrupling of the exponent in
the scaling of $p_{{\cal W}\neq 0}$ (reflected in the frequency of trapping $|{\cal W}|=1$) with quench rate from $\frac 1 8$ predicted originally \cite{Zurek85} for large typical $\cal W$ 
to $\simeq \frac 1 2$ 
was reported in an impressive sequence of experiments that involved rapid cooling of tunnel Josephson Junctions \cite{RM1,Quad,Quadtoo} where $|{\cal W}| \ll 1$. 
The authors (who expected doubling from $\frac 1 8$ to $\frac 1 4$, and regarded their data as inconsistent with KZM) start Section VI ``Theory'' in Ref. \cite{Quad} with the statement: {\it ``The value of ... 0.5} [obtained from the fit 
of the frequency of detections of $|{\cal W}|=1$] {\it is obviously in disagreement with our earlier prediction of ... 0.25 given in} \cite{KMR}...'' They go on to attribute the discrepancy with Ref. \cite{KMR} to possible fabrication flaws resulting in the `proximity effect' \cite{RM1}.

The ``{\it ...0.25 given in} ... \cite{KMR}...'' is twice $\frac 1 8 $, the exponent of the scaling of the typical winding number (either dispersion $\sqrt {\langle {\cal W}^2 \rangle }$, or the average $\langle | {\cal W} | \rangle$) for large $\cal W$ with the quench rate for the relevant $\nu= \frac 1 2, \ \nu z =1$ \cite{Zurek85,DSZ12}. The theory put forward here shows that the observed \cite{RM1,Quad,Quadtoo} quadrupling of the exponent of the trapping frequency to $ \frac 1 2$
(square root of the quench rate ${\tau_Q}^{-1}$) actually verifies KZM predictions.

\begin{figure}
\begin{center}
\includegraphics[width=0.95 \linewidth]{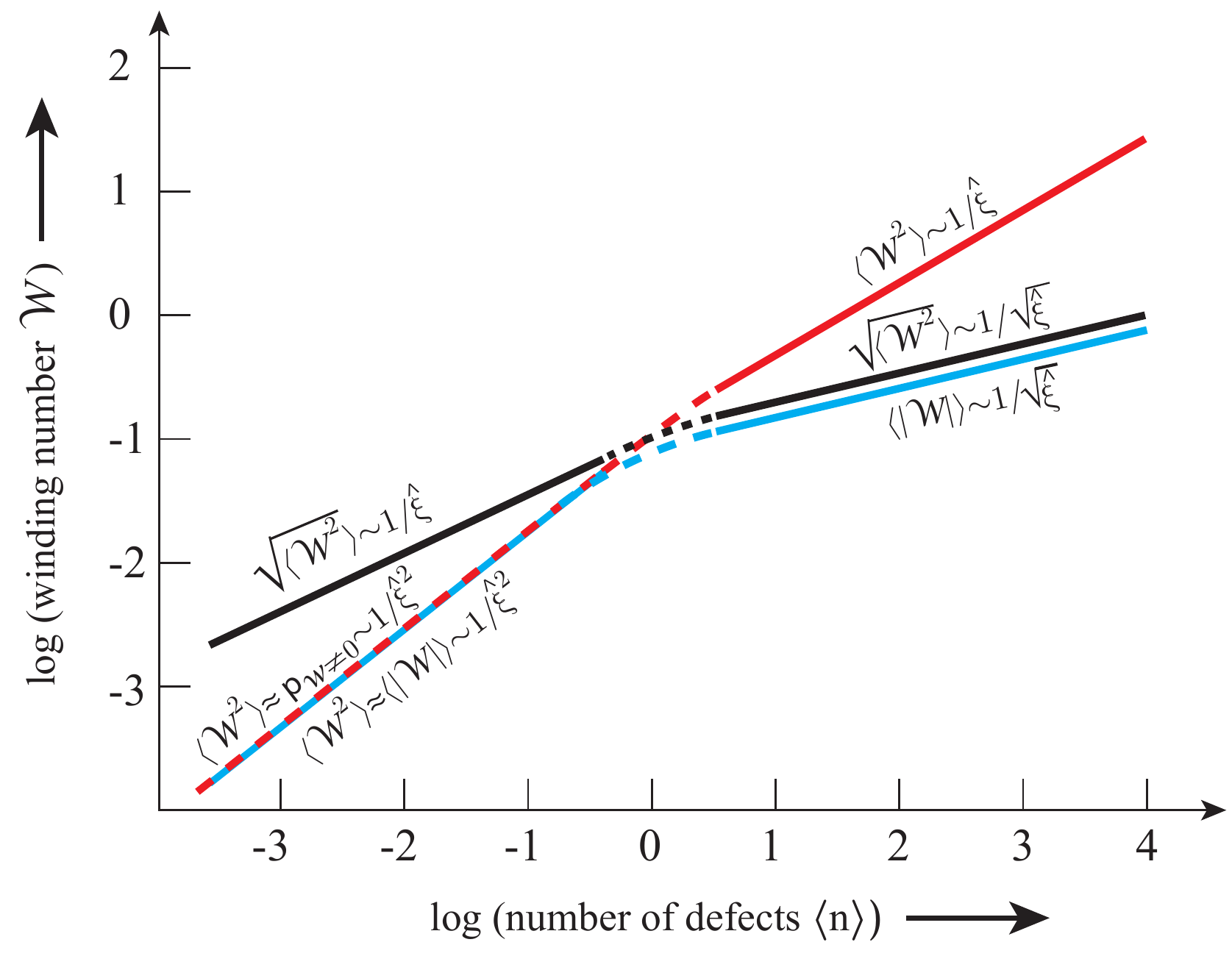}
\end{center}
\caption{\label{tilt} 
Schematic illustration of the changes of the tilt of the scaling of dispersion, $\sqrt {\langle {\cal W}^2 \rangle }$, its square, and the average $\langle | {\cal W} | \rangle$ of the winding numbers with the expected number of defects trapped inside $\cal C$. For loops that trap many defects, $\langle n \rangle \gg 1$, dispersion and the average absolute value of $\cal W$ scale similarly, 
$\sqrt {\langle {\cal W}^2 \rangle} \sim | {\cal W} | \sim \sqrt {{\cal C}  /{\hat \xi} }$, Eqs. (1) and (10a). 
However, different tilts corresponding to exponents that control slopes of dispersion and average absolute value of the winding number occur as $\langle n \rangle \ll 1$. Compared to $ \sqrt {{\cal C}  /{\hat \xi} }$,  the slope of dispersion {\it doubles},  $\sqrt{\langle {\cal W}^2 \rangle }  \sim {\sqrt A_{\cal C} / {\hat \xi}}$ while the slope of the average absolute value {\it quadruples} so that $| {\cal W}| \simeq p_{|W| = 1} \sim A_{\cal C} / {\hat \xi}^2 \sim {\langle {\cal W}^2 \rangle }  $ when $\langle n \rangle \ll 1$, where $A_{\cal C}$ is the area enclosed by the contour ${\cal C}$. (Note that $\langle n \rangle = A_{\cal C} / {\hat \xi}^2$).
}
\end{figure}


Our conclusions about the tilt of the scaling  between the two regimes should not be applied uncritically: We have deduced tilt using a specific model -- defect - antidefect pairs. Nevertheless, key steps that led to the reevaluation of scaling are statistical in nature, and therefore predicted tilt of the scalings is reasonably independent of the details. For instance, our conclusion about scaling in the $\langle |{\cal W}| \rangle \ll 1$ case, Eq. (11), (or Eq. (1')) does not (in contrast to Eq. (1)) depend on the random distribution of vortex pairs: It would equally well apply to random distribution of vortex charges (even though the latter gives incorrect predictions in the $\langle |{\cal W}| \rangle \gg 1$ regime). Still, while the predictions seem more robust than the model we have started with, it is possible that other effects may influence scalings observed in experiments or in computer simulations of specific physical systems used to test KZM\footnote{For example, energetics overrides simple statistics in an example that does not involve winding: Quantum Ising in a transverse field exhibits exponential probability of appearance of a single defect that cannot be captured by any of the scalings listed above (see e.g. \cite{ZurekZoller,Dziarmaga}) and, yet, that is tied to KZM via Landau-Zener theory \cite{Damski}. }. 

It would be intriguing to compare predictions of tilt with experiments where both regimes $ \langle{\cal W}^2\rangle \gg 1 $ and $ \langle{\cal W}^2\rangle \ll 1 $ could be explored in the same system. Multiferroics (see e.g. Ref. \cite{Cheong}) may be an attractive area for such research, providing that the KZM is indeed applicable, as argued in Ref. \cite{Spaldin}.

\section{Why pairs, and how random?}

Why should topological defects come in vortex-antivortex pairs? In general, I know of no really fundamental reason for that. Of course, it is the broken symmetry -- e.g. the phase of the condensate wavefunction --  which is ultimately random. This results in anticorrelations of the topological charges \cite{LM} corroborated by the experiment \cite{Golubchik}. However, anticorrelation does not necessarily imply pairing we have assumed. So, in effect, ultimately the pairing picture does not always have a firm physical basis. It just happens to be a convenient way of summing up correlations that are there in the wake of the transition. Its biggest advantage is a simple way to justify Eq. (1). It also provided stimulus to arrive at scaling tilts, Fig. 2, although our conclusions about tilts and power law exponents seem to be (within reason) independent of the assumption of pairing.

Indeed, arbitrary pairing may not be enough to justify our conclusions. Our discussion above assumes that the sizes $l$ of pairs are not too large -- e.g., not much larger than a typical distance $L$ deduced from the density $L^{-2}$ of defects. This seems to be the case in some situations (e.g., phase transitions in multiferroics, see Ref. \cite{Cheong}). If this assumption were to fail, our reasoning that led to Eq. (1) might also fail. For instance, when gauge fields dominate defect charges are correlated \cite{HindmarshRajantie}, with clusters of vortices of same charge \cite{SBZ}. One can still imagine defect-antidefect pairs, but now $l$ could be large compared to $L$.

In spite of the above admission (that pairing we have postulated was really a ruse to get Eq. (1) from a simple intuitively appealing picture of a distribution of defects rather than phases, and to gain insight into tilts of the scalings of winding numbers around $|{\cal W}| \sim 1$), there are clearly examples where one can expect pairing on physical grounds. For instance, in superfluids and superconductors one can use a 3-D random walk on a lattice as a model for the emerging vortex line. Such random walks seem to have many features of the vortex lines seen in computer simulations (of, e.g., cosmic strings \cite{VV}). Self-avoiding random walk naturally defines vortex-antivortex pairs of interest to our discussion while piercing a 2-D plane -- quite simply, consecutive piercings have opposite topological charge. Therefore, one can regard them as vortex-antivortex pairs. This pairing would be reasonably unambiguous for $\sim 30\%$ of the total length of 3-D vortex line that comes in loops (as is estimated in computer simulations and consistent with Polya's theorem on random walks \cite{Polya}). The other $\sim 70\%$ of the random walk is in an infinite string. 
When consecutive piercing are paired up it is conceivable the some of them may be quite far apart. Still, it seems plausible that such infinite random walk (once it passes through a 2-D plane) will likely pass through it again nearby, ``on its way back''. Thus, there is a possible way to at least motivate pairs.

We also note that there are parallels between the (gist of the) estimate of the winding number given above and the (simpler, and rightly famous) ``Buffon's needle problem'', paradigmatic illustration of the geometrical approach to probability \cite{Gnedenko}. 
In Buffon's problem a needle of fixed length $l$ is tossed at random on a plane that is ruled with a series of parallel lines that are a distance $a$ apart (sort of like a floor). The object is to determine the probability that the needle will intersect one of the lines. The probability turns out to be $l / (\pi a)$. This result can be even used to ``experimentally'' estimate the value of $\pi$. (One might regard this as a precursor to Monte Carlo approach.)

In our case, needles are vortex pairs, and instead of a floor ruled with parallel lines we have a single circumference which we assume is a circle. Moreover, our needles are ``polarized'', but we assume that this polarization is random, which eventually leads to Eq. (1). The version of Buffon's needle problem with the needle dropped on a collection of concentric circles (sort of like a shooting target) has been also discussed \cite{Khamis}, but only recently, and the solution of the problem involves fairly painful integration and assumptions (such as many concentric circles with radii $ r, ~2r, ~3r$, etc.) we could not justify. 
The basic conclusion (in the case when it is useful for us) confirms what was easy to guess: That the number of needles that fall onto ${\cal C} \gg l$ is proportional to $\cal C$.


\section{Discussion, summary and outlook}

We have seen that the scaling of the winding number subtended by a circle of circumference $\cal C$ from a collection of vortex-antivortex pairs is given by Eq. (1). That is, it is consistent with KZM when the typical winding numbers are large. Moreover, we have predicted -- based largely on statistical considerations -- how the dependence of the winding number on the number of topological defects inside $\cal C$ will change when loops are so small that it is unlikely they can trap more than one defect inside. This prediction could be tested in future experiments (e.g., in multiferroics  \cite{Cheong}) and may also help reconcile outcomes of past experiments \cite{RM1,Quad,Quadtoo} with KZM.

The basis of our discussion of the tilt of scalings between loops that trap many defects vs. loops that trap at most a single defect is a straightforward application of the (geometric) probability theory. Scalings are dominated by the random walk behaviour of the typical values of $\cal W$ for large winding numbers, where $\sqrt{\langle {\cal W}^2 \rangle } \sim \langle |{\cal W}| \rangle$. However, when the only likely values of $|{\cal W}|$ are 0 and 1, probability of non-zero $\cal W$ mandates $\sqrt{\langle {\cal W}^2 \rangle }\sim \sqrt {p_{{\cal W}\neq 0}} $, behavior that differs from the average absolute values $\langle |{\cal W}| \rangle \sim p_{{\cal W}\neq 0}$. 

The intuitive argument for generalization invokes a loop $\cal C$ ``lassoing'' topological defects as it is dropped, at random locations, on a plane with fluctuating configurations of the order parameter that result in defects: A loop $\cal C$ that becomes superconducting ``lassoes'' flux lines arising from random configurations of the order parameter inside $\cal C$, thus trapping W. Therefore, instead of sampling random configurations of the order parameter inside a fixed loop we are sampling them on an infinite plane (where all the possible configurations occur) by dropping $\cal C$ at random locations. 

The change, by a factor of four, in the exponent of the power law dependence between scaling of typical values of the winding number for large $\cal W$ vs. the frequency of detection of ${\cal W} \neq 0$ when $\langle | {\cal W}  | \rangle$ is fractional is a product of two doublings, or, rather of two independent reasons to take a square root. 

The first square root is specific to the definition of $\cal W$. Winding number is a consequence of an accumulation of the net phase difference. It is proportional to the {\it square root} of the number of steps. When $\cal C$ is so small so that at most one step can be taken, there is no need to take a square root, as $\sqrt 1 = 1$. This doubling of the power law exponent 
would not arise if the observable in the experiment was the number of changes of the phase irrespective of the sign (e.g., if each phase change -- each vortex-antivortex pair intercepted by $\cal C$ -- contributed a ``kink'', and if the number of such kinks rather than $\cal W$ was of interest). 

The second square root arises from the change in the estimate of the number of pairs that intersect contour $\cal C$ as its circumference shrinks. For large ${\cal C} \gg \hat \xi$ the number of pairs that intersect $\cal C$ matters. That number is proportional to the product of $\cal C$, $ \hat \xi$, and pair density $\sim \hat \xi^{-2}$, so in the end it scales as ${\cal C} / \hat \xi$. Pairs that do not intercept $\cal C$ -- that have both members inside -- do not contribute, so the area $A_{\cal C}$ subtended by $\cal C$ does not matter when it is large compared to ${\cal C} \hat \xi$. However, when ${\cal C} < \hat \xi$, only one member of a pair will be typically lassoed by $\cal C$, with the probability that is proportional to $A_{\cal C}$ times the pair density $\sim \hat \xi^{-2}$, so it scales as $A_{\cal C}/\hat \xi^{2}$. The exponent of the power law dependence on $ \hat \xi$ (and, hence on the quench time $\tau_Q$) doubles as a result. 

As we have already hinted, there may be cases where only one of these two doublings matters. For instance, if each step of phase could be regarded as a kink on $\cal C$, and kinks did not annihilate (so their total number rather than the net accumulated phase could be measured), the doubling arising from the random walk (and relevant for winding numbers) would not be there, but the doubling arising from the transition from the regime where the number of pairs is proportional to ${\cal C}/\hat \xi$ to where it behaves as $A_{\cal C}/\hat \xi^{2}$ would be still expected. Therefore, frequency of trapping of a single kink would follow a power law with an exponent that is twice the exponent for large kink numbers. It remains to be seen whether such doubling can be held responsible for the scaling of defect frequency in buckling of inhomogeneous Coulomb crystals in ion traps \cite{Pyka,Ulm,Haljan}.


Our heuristic picture -- $\cal C$ lassoing vortices -- may not be generally applicable. Specific physics (of superconducting loops, tunnel Josephson Junctions, BEC's, etc.) could change  aspects of behaviour that affect winding numbers trapped inside $\cal C$ (as suggested by \cite{NZ,RM2,DTZ}). Using the above ``lasso'' analogy one could say that presence of $\cal C$ may constrain or modify fluctuations in its vicinity, and, therefore, alter the nature of the process responsible for formation of defects. For instance, defects may have more time to ``run away'' from where they get trapped by $\cal C$ when the quench is slower.

Thus, while statistical nature of our discussion suggest that its conclusions e.g., about the change of slopes (summed up in Fig. 2) may generalize beyond the model used to justify them, caution in extrapolating it is essential. One conclusion that aspires to generality is the need to treat small systems, with sizes comparable to $\hat \xi$, with care.

The other conclusion is that while, for large $\cal W$, both the dispersion and the average absolute value can be regarded as reasonable estimates of the ``typical magnitude'' of the winding number, this is no longer so when the only likely values of $|{\cal W}|$ are 0 and 1. In a sense, this is really no surprise: A typical value of $\cal W$ in this case is, quite simply, 0. However, a ``typical deviation from ${\cal W}=0$'' behaves differently -- scales differently with the size of the loop -- depending on whether it is represented by $\sqrt{\langle {\cal W}^2 \rangle }$ or by $|{\cal W}|$.

It would be also interesting to investigate experimentally what happens when $\cal C$ is deformed. KZM predicts that the scaling of Eq. (1) should remain a good approximation at least as long as the contour does not have a structure on scales comparable or smaller than $\hat \xi$ (which we can think of as an estimate of the pair size $l$). Thus, when the loop retains its circumference, but changes its shape (within reason -- see discussion in Section 3) Eq. (1) should continue to hold. For example, imagine an elongated stadium with the circumference $\cal C$, and with the two straight sides of length $D$ separated by $2R$ -- the diameter of the semicircles -- so that the circumference $2D+2\pi R$ can change while the area $DR + \pi R^2$ remains the same. Equation (1) is still expected to hold. Presumably, only when the width of the stadium $2R$ falls below $l$ -- the typical separation of pair members -- the winding number would begin to fall below the estimate of Eq. (1) since they have to decrease (as, eventually, in the limit of $2R\rightarrow 0$, a stadium-shaped lasso loop can trap no defects, so that the winding numbers must also vanish).

Extension of our discussion to monopoles in 3-D would be also intriguing. We expect that the basic ideas behind the argument will apply. Thus, only pairs that are bisected by the surface $\cal S$ (that now plays the role analogous to the contour $\cal C$) can contribute to the  topological charge of the volume enclosed by $\cal S$. Therefore net topological charge trapped inside the surface area $\cal S$ should approximately scale with the surface area as $ \sqrt {\cal S} \sim R$, and not as $\sim \sqrt {R^3} = R^{\frac 3 2 }$, i.e., with the square root of the volume it encloses. This assumes $R \gg l$ and the caveats concerning deformations of $\cal S$ analogous to these illustrated by the 2-D stadium example discussed above and in Section 3; we also exclude peninsula-like protrusions, isthmus, and other similar deformations that might results in flattening of $\cal S$ so that the typical size $l$ of defect pairs is small compared to the thickness of the volume subtended inside). 


\ack

Discussions with Sang-Wook Cheong, Adolfo del Campo, Jacek Dziarmaga, Valery Kiryukhin, Raymond Rivers, and Vivien Zapf are gratefully acknowledged. This work was supported by the Department of Energy through the LDRD program in Los Alamos.

\end{document}